\documentstyle[twocolumn,amssymb,aps,psfig,float]{revtex}
\begin{document}

\baselineskip11pt
\newcommand{\bsigma}{\mbox{\boldmath $\sigma$ } }
\newcommand{\bS}{\mbox{\boldmath $S$ } }
\newcommand{\bT}{\mbox{\boldmath $T$ } }
\newcommand{\bLambda}{\mbox{\boldmath $\Lambda$ } }
\newcommand{\bmu}{\mbox{\boldmath $\mu$ } }
\newcommand{\bk}{\mbox{\boldmath $k$ } }
\newcommand{\bR}{\mbox{\boldmath $R$ } }

\twocolumn[\hsize\textwidth\columnwidth\hsize\csname @twocolumnfalse\endcsname

\title{ 
Orbitally Degenerate Spin-1 Model for Insulating V$_2$O$_3$
}
\vskip0.5truecm 
\author{F. Mila$^{(a)}$, R. Shiina$^{(b)}$\cite{*}, F.-C. Zhang$^{(c)}$, A.
Joshi$^{(c)}$, M. Ma$^{(c)}$, V. Anisimov$^{(d)}$, and T. M. Rice$^{(b)}$}

\address{
$^{(a)}$ Laboratoire de Physique 
Quantique, Universit\'e Paul Sabatier, 31062 
Toulouse, France\\
$^{(b)}$ Theoretische Physik, ETH-H\"onggerberg, CH-8093 Z\"urich, Switzerland\\
$^{(c)}$ Department of Physics, University of Cincinnati, Cincinnati, 
Ohio 45221\\
$^{(d)}$ Institute of Metal Physics, Russian Academy of Sciences, 620219,
Ekaterinburg, GSP-170, Russia }
\vskip0.5truecm
      
\maketitle

\begin{abstract}
\begin{center} 
\parbox{14cm}
{
Motivated by recent neutron, X-ray absorption and resonant scattering
experiments, we revisit the electronic structure of V$_2$O$_3$. We propose a
model in which S=1 V$^{3+}$ ions are coupled in the vertical V-V pairs 
forming two-fold
orbitally degenerate configurations with S=2. Ferro-orbital ordering of the V-V
pairs gives a description which 
is consistent with all experiments in the antiferromagnetic insulating phase.
}
\end{center}
\end{abstract}
 
\vskip2pc
]

Although the metal-insulator transition in V$_2$O$_3$ has long been
studied as a classic Mott-transition\cite{McWhan1,McWhan2,mott}, 
the detailed electronic
structure remains open. Recently new experimental techniques have been 
applied but these have not resolved the issue. Rather they have
reopened the long standing controversy between an $S=1$ model without
an orbital degeneracy and the $S=1/2$ orbitally degenerate model of
Castellani et al.\cite{castellani}. In this Letter we propose a new 
model for the
AF-ordered insulating (AFI) phase based on the molecular orbitals of
the $c$-axis V-V pairs, which combines features of both existing
models and which reconciles the apparently conflicting experiments
supporting each.

\begin{figure}[hp]
\centerline{\psfig{figure=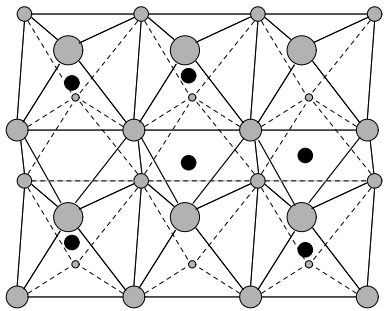,width=5.0cm,angle=0}}
\vspace{0.5cm}
\caption{Corundum structure of V$_2$O$_3$. The V-ions (solid circles) are
arranged in V-V pairs along the c-axis (face-sharing octahedra) and a honeycomb
lattice in the other directions (edge-sharing octahedra).}
\label{fig1}
\end{figure}

The V-ions in the corundum structure of V$_2$O$_3$ sit in a
O-octahedron with a small trigonal distortion causing a small
splitting in the non-bonding $t_{2g}$-shell between the
$a_{1g}$-orbital oriented along the $c$-axis and doublet planar
$e_g$-orbitals (see Fig. 1). In their early work, Castellani et al. proposed 
that
one electron of the $3d^2$ V$^{3+}$-ion entered a spin singlet
covalent $a_{1g}$-bond in the V-V pair while the remaining electron
was in the $e_g$-doublet. Orbital ordering of these $e_g$-doublets
allowed them to explain the unusual magnetic structure of the
AFI-phase with inequivalent $n.n.$ exchange constants in the $a-b$
plane (2 antiferromagnetic, 1 ferromagnetic)\cite{Moon,Word,Bao}. Paolasini et
al.\cite{paolasini} interpreted their recent resonant x-ray experiments as a
confirmation of this orbital ordering. On the other hand the polarized 
soft x-ray experiments by Park et al.\cite{park} showed a coexistence of both
$(e_g\,e_g)$ and $(e_g\,a_{1g})$ configurations in roughly equal
amounts and these led Ezhov et al.\cite{ezhov} to argue for a $S=1$ 
model with a
$(e_g\,e_g)$ configuration and no orbital degeneracy. This is favored
by the atomic Hund's Rule whose strength, as they point out, is not
screened in the crystal. The differing planar exchange constants they
attribute to the monoclinic distortion in the AFI-phase. 
Yet general
considerations of the phase diagram\cite{rice} and NMR
investigations\cite{takigawa} all point
towards to the presence of an orbital degeneracy.

Here we take a different approach to the AFI-phase and start from an
atomic limit but consider first the V-V pairs, since the intersite
$a_{1g}$-hopping matrix elements are the largest\cite{castellani}.
Keeping a strong Hund's Rule coupling, as proposed by 
Ezhov et al., leads us to molecular orbitals for a V-V pair consisting
of a superposition of $(e_g\,e_g)$ on one V-site and of
$(e_g\,a_{1g})$ on the second site with a total spin $S=2$. This
delocalized molecular orbital has also a two-fold degeneracy due to a
choice in $(e_g\,a_{1g})$ among the $e_g$-doublet. Next we consider
planar hopping processes and show that in a reasonable parameter range 
the real spin (RS) structure is the most stable. This state has a
ferro-arrangement  of the molecular orbitals which agrees with the
monoclinic structure and, as we shall see, also with the x-ray experiments of
Paolasini et al. 

Let us start with a description of a vertical pair. Following
Ref.\cite{castellani}, the two $e_g$ orbitals\cite{com1} are specified by a 
further index
as $|e_{g1} \rangle = 1/\sqrt{2}(|d_{yz} \rangle - |d_{zx} \rangle )$
and $|e_{g2}\rangle 
= 1/\sqrt{6}(2 |d_{xy} \rangle - |d_{yz} \rangle - |d_{zx} \rangle )$, while the
$a_{1g}$ orbital is given by $|a_{1g} \rangle 
= 1/\sqrt{3}( |d_{xy} \rangle + |d_{yz} \rangle + |d_{zx} \rangle )$.
For each V ion, the d-orbitals are defined in a local coordinate system
whose axis point towards the surrounding O ions, and thus refer to
different, symmetry related orbitals for the different V ions in the
unit cell. Consequently, the $e_{g1}$ and $e_{g2}$ orbitals on the
two V ions of a vertical pair are not identical. This will be important
when we compare our results to that of resonant scattering experiments.
The intra-atomic interaction is described by three parameters: 
$U$, the Coulomb interaction in the same orbital, 
$U'$, the Coulomb interaction in different orbitals, and $J$, 
the Hund's Rule coupling, which we assume satisfy
the usual relation for $t_{2g}$ orbitals: $U=U'+2J$. The trigonal crystal field
induces an energy splitting $\Delta$ between the low-lying $e_g$ orbitals and
the excited $a_{1g}$. Finally, the hopping integrals are denoted by
$t^\delta_{ij}$ where $\delta=a,b,c,d$ stands for the direction of the 
bond ($a,b,c$: bonds inside the hexagonal planes, $d$: vertical
bond) while $(i,j)=1,2,3$ denote the orbitals ($e_{g1}$,$e_{g2}$ and
$a_{1g}$ respectively). 

The main difference with Ref.\cite{castellani} comes from the values 
of the
interaction parameters. The values used in Ref.\cite{castellani} ($U\simeq 2
eV$, $J=0.2 eV$) are now believed to be much too small: Recent estimates based
on LDA+U calculations\cite{ezhov} are in the range $U\simeq 5$ eV and 
$J\simeq1$ eV. 
It turns out that this makes a dramatic difference for the ground state of a
V-V pair. To be specific, if we consider the same hopping and crystal
field parameters as in Ref.\cite{castellani}, and if we fix the ratio $J/U=0.1$
to the value they used, 
there is a level crossing as a function of $U$ between two very different 
situations. At small $U$, the ground state is 3-fold degenerate, with
3 levels nearby. This corresponds to the limit of Ref.\cite{castellani} where 
two electrons go
into the bonding molecular orbital built out of $a_{1g}$ orbitals, the other
two electrons being described by a spin 1/2 - pseudo spin 1/2 Kugel-Khomskii 
model\cite{kugel}. At large $U$ however, the ground state is 10-fold degenerate. It
corresponds to a total spin 2 with a two-fold degenerate orbital state. 
Since by symmetry $t^d_{ij}=0$ if $i\ne j$, this orbital wavefunction 
can actually be written down explicitly: 
\begin{equation} 
 | \pm \rangle = \frac{ 
 |e_g, a_{1g} \rangle \otimes |e_{g1}, e_{g2} \rangle
+ |e_{g1}, e_{g2} \rangle \otimes |e_g, a_{1g} \rangle}{\sqrt{2}} 
\label{pm}
\end{equation}
where $e_g$ stands for $e_{g1}$ ($e_{g2}$) in $| - \rangle$ ($| + \rangle$). 
This situation is generic for a large range of $J/U$ including $J/U=0.2$, 
and with the parameters proposed in Ref.\cite{ezhov}, we found that 
the ground state is clearly of this second type.

It is interesting to compare this state with the spin 1 picture of Ezhov et al.. 
When the Hund's Rule coupling is large, all low-lying states can indeed be
described by considering only the states with total spin 1 at each site.
However, the resulting effective Hamiltonian for a pair of sites is not 
simply a
Heisenberg Hamiltonian, since this would correspond to only 9 low-lying states.
In fact there are 81 low-lying states, suggesting that it one wants to describe
this system with a spin 1 operator, $\vec S$, at each site, one should also 
include a pseudo-spin 1 operator, $\vec T$, to describe the quasi-degeneracy 
of the $t_{2g}$ orbitals. This orbital degree of freedom is 
crucial since it is responsible for a factor 2 in the 10-fold 
groundstate degeneracy.

These results suggest that, instead of starting from a spin-orbital model 
with a spin 1/2 and a pseudo-spin 1/2 at each V site, one should start from a
spin-orbital model in which each vertical V-V pair is decribed by a spin 2
for the total spin, say $\vec \sigma$, and a pseudo-spin 1/2 for the
orbital degeneracy, say $\vec \tau$, $\tau^z=1/2$ ($\tau^z=-1/2$)
corresponding to $| + \rangle$ ($| - \rangle$) in Eq.\ref{pm}. The low energy
properties are then determined by the way the degeneracy is lifted when these
pairs are coupled by the in-plane hopping integrals. Since these hopping
parameters are small, we can treat them within second-order perturbation theory.
For simplicity, we include only the largest hopping integral $t^a_{23}\equiv t$,
and the corresponding hopping integrals for directions $b$ and $c$, in the
present discussion. We have checked that the conclusions are unaffected by this
simplification. The second order effective spin-orbital Hamiltonian for n.n. 
along the $a$-axis then reads\cite{com2}:
\begin{equation}
\tilde{H}_{\rm eff}(a) = G \vec \sigma_l \cdot \vec \sigma_m 
+  \frac{1}{4} G_3 (\tau_l^z+\tau_m^z) \vec \sigma_l \cdot \vec \sigma_m, 
\end{equation}
with 
\begin{eqnarray} 
G= -\frac{1}{3}G_1 + \frac{1}{3}G_2 
         + \frac{3}{4}G_3\ \  &,& \ \ 
G_1 = \frac{ t^2}{4( U'- J )} \nonumber \\
G_2 = \frac{ t^2}{4( U'+ 2J )}\ \  &,\ \ &
G_3 = \frac{ t^2}{4( U + J \big)}  
\end{eqnarray}
The effective Hamiltonians for n.n. along the $b$- and $c$-axes are easily 
obtained by the trigonal 
rotation of $\tilde{H}_{eff}(a)$ equivalent to the following replacement 
of the orbital pseudo-spin 
$ \tau^z \rightarrow -1/2 \tau^z \pm \sqrt{3}/2 \tau^x $. 
While there is a strong anisotropy in orbital space, 
the interaction preserves $SU(2)$ symmetry for the spin operator 
$\vec \sigma$.

Remarkably enough, the symmetry properties of this model 
are quite similar to the Kugel-Khomskii model for the cubic 
perovskite\cite{kugel}. In fact one can regard the corundum lattice 
as a distorted simple cubic ($sc$) lattice of the V-V pairs. 
This analogy is useful to give a systematic analysis 
for such a complicated system. Namely, it is promising that 
the stable magnetic phases of Kugel-Khomskii-type models
are collinear two-sublattice orderings with associated orbital orderings. 
Within this criterion the possibility is naturally restricted 
into G, F, C, and A-type magnetic patterns\cite{shiina}. 
In this language, the realistic magnetic structure of V$_2$O$_3$ corresponds 
to the $C$-type arrangement in the pair $sc$ lattice: 
One of 3 in-plane bonds is ferromagnetic and other two bonds are 
antiferromagnetic.

\begin{figure}[hp]
\centerline{\psfig{figure=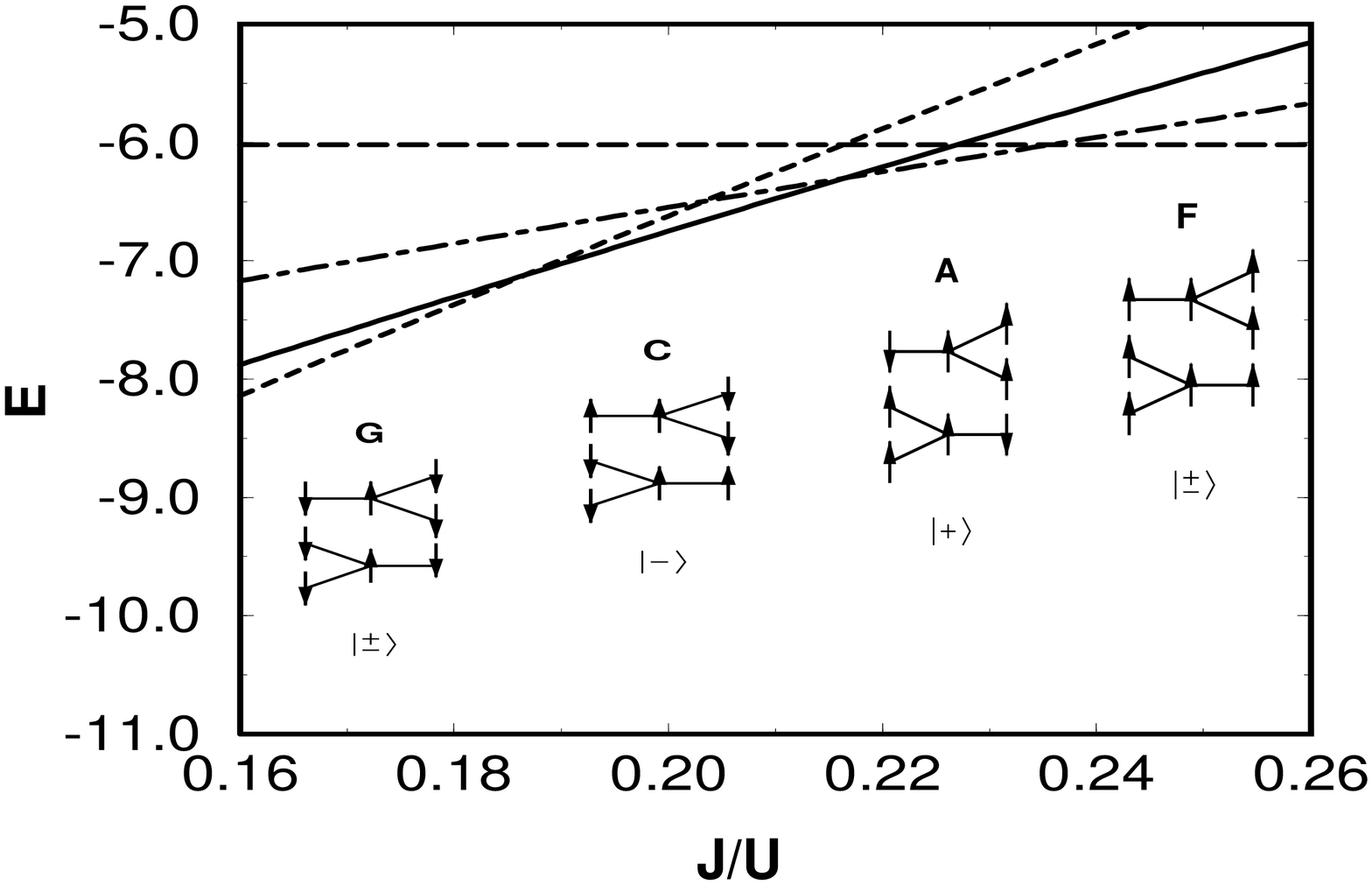,width=9.0cm,angle=0}}
\vspace{0.5cm}
\caption{Energy comparison of various magnetic patterns. The letters refer to
the notations of Ref.{\protect \cite{shiina}} for the $sc$ lattice. 
The pictures give the
corresponding magnetic pattern for the corundum lattice. The symbols
$|\pm \rangle $, $|+ \rangle $, and $|- \rangle $ indicate which of the
molecular orbitals of Eq. \ref{pm} is consistent with 
the magnetic pattern.
The orbital degeneracy is lifted for states C and A, but not for states
G and F.}
\label{fig2}
\end{figure}

Keeping in mind these relations, we have examined the stable phase 
in the molecular model by comparing the energies of all magnetic 
phases. This has been done, as for the Kugel-Khomskii model, within 
a mean-field
decoupling based on the order parameters $\langle \tau^\alpha \rangle$,
$\langle \sigma^\alpha \rangle$, and $\langle \tau^\alpha \sigma^\beta \rangle$.
Details will be given in a forthcoming paper\cite{shiina2}.
The results are plotted in Fig.\ref{fig2} as a function of Hund's Rule 
coupling $J$. It turns out that the stable phase changes successively
from G to F phase as $J$ increases. 
In order to gain energy by the orbital-dependent $G_3$ term, 
the symmetry-broken C and A phases are stabilized 
in the intermediate-$J$ region. 
In particular the realistic $C$-type phase is found to be the lowest 
for $ J/U $ around $ 0.2 $, which agrees with the estimates of Ezhov et al., 
and 
which is consistent also with the stability region for the $S=2$ degenerate 
molecular orbitals of a V-pair. For this phase, the orbital order parameter is
ferromagnetic with $\tau^z<0$, i.e. the $e_{g1}$ orbital is favoured (state
$|-\rangle $ of Eq. \ref{pm}).
Such a ferro molecular orbital order will
cause an effective uniaxial stress on the lattice degrees of freedom, leading 
to a uniform rotation of V-V pairs. This is consistent 
with the monoclinic distortion proposed by 
Dernier and Marezio\cite{dernier}. 

The physical picture that emerges from this model is very encouraging.
First of all, the observed magnetic arrangement\cite{Moon,Word,Bao} 
is consistent with this model
for reasonable values of the parameters. Second, the atomic configuration is
a mixture of ($e_g e_g$) and ($a_{1g} e_g$), in agreement with X-ray
absorption\cite{park}. 
Third, there is an orbital degree of freedom whose ordering is consistent with
the monoclinic distortion of the low-temperature phase\cite{dernier}. 
It corresponds to
choosing between $e_{g1}$ and $e_{g2}$ for the V-V pairs.

The results of our  model are also consistent with the resonant x-ray
scattering experiment of Paolasini et al.\cite{paolasini}. In that experiment,
resonant scattering was observed in the low temperature phase at wavevector
$q=(111)$ and at energies corresponding to the transition from 1s to unoccupied
3d states on V-ions. Since this is
forbidden by symmetry if all the orbital configurations on V-atoms are equal,
this led Paolasini et al. \cite{paolasini} to conclude to the existence of
orbital ordering.  They further suggested that their experiment can be taken
to confirm one of the orbital ordered phases previously proposed by Castellani
et al\cite{castellani}.  We now show that the orbital ordering proposed in this
paper for our model, while different from that interpreted by Paolasini et
al., is also consistent with the experimental results.

The resonant  scattering amplitude as a function of the energy
$\omega$ and  $q$ of x-rays for a crystal $V_2O_3$ is
given by  
$F=\sum_{i=1}^{8}
e^{i\vec{q}\cdot\vec{\rho_i}}f_i (\omega)$, 
where $f_i (\omega)$
is the amplitude contributed from the V atom at position $\vec{\rho_i}$ in the
monoclinic unit cell of $V_2O_3$. The low-temperature monoclinic lattice of $V_2O_3$
has eight atoms in a unit cell (Fig.3). Atoms 1-4  have
spin-up magnetic moments and atoms 5-8  have spin-down
magnetic moments. F at $q=(111)$ is given by
$$F_{111}=(f_1-f_5+f_8-f_4)e^{i \alpha}+ (f_2-f_6+f_7-f_3)e^{-i
\alpha}$$
where $\alpha $ is a phase factor which depends on $\vec \rho_1 - \vec \rho_2$.
$f_i(\omega)$ depends in general on the magnetic moment
and orbital occupation\cite{hannon}. 
The nonvanishing intensity of (111) reflection for this energy implies that the
combinations $(f_1-f_5+f_8-f_4)$ and $(f_2-f_6+f_7-f_3)$ are nonzero.  The
ferro-orbital phase in our model exhibits this feature for the following 
reasons. As discussed before, the $e_g$ orbitals are defined with
respect to a local co-ordinate system on each V ion. In particular, for the
two V ions on a vertical bond, they are related  by a
rotation around the $y$-axis:  $C_2(x,y,z)=(-x,y,-z)$ (the trigonal coordinate
system is used here with $z$-axis directed perpendicular to the hexagonal
plane) while for the V ions in the same hexagonal plane local coordinate
systems are identical.
Thus, the ferro-orbital phase actually corresponds to having
different orbitals on alternate hexagonal planes. Denoting these as 1
and 2 and denoting $u$ and $d$ for the spin up and down magnetic state, we
then have e.g. 
$$f_1-f_5+f_8-f_4= f(1,u)-f(1,d)+f(2,d)-f(2,u)$$
Since f depends on both the orbital occupation  and the magnetic moment of
the V-atom, $f(u) \neq f(d)$. Thus, our model gives nonzero $F_{111}$, and is
qualitatively consistent with the experimental observation of Paolasini et al.
\cite{paolasini}. More work is needed to compare our theory with the observed
polarization and the azimuthal dependences of the resonance intensity.

According to this explanation, the intensity of the (111) reflection 
is not simply
a direct consequence of the orbital order, but comes both from magnetic and 
orbital order. This should be contrasted to 
Paolasini et al.'s explanation based on Ref.\cite{castellani}, 
where the form factor of Eq. (5) is non zero 
because the orbitals occupied on the two V of a vertical pair 
are different linear combinations of $eg_1$
and $eg_2$. While Castellani et al.'s picture is very 
specific to their S=1/2 model, there is room {\it a priori} 
within the S=1 model for 
a similar orbital ordering. Energetic considerations then show that the only
serious candidate is an orbital ordering in
which one V of a vertical pair would be in the  ($eg$ $a_{1g}$) configuration,
while the other one would be in the ($eg$ $eg$) configuration.
To be more precise, the model considered thus far corresponds to having all
interplane hopping 
amplitudes much larger than intraplane ones. If we maintain the 
limit of $t_{33}^d$ large, but allows $t_{11}^d$ and $t_{22}^d$ to be
comparable to intraplane hoppings, it can be shown that the problem 
can then be mapped into a transverse field Ising model.  Details will be
given in a forthcoming publication~\cite{joshi}.  Here we briefly summarize
the main results. In this mapping, there is an Ising spin associated with
each V-V pair which corresponds to the orbital occupation of $(e_g e_g)$ 
and $(e_g a_{1g})$ on the two site. The transverse field strength $h$
is given by the energy difference of the bonding and antibonding states of a
V-V pair, a measure of the orbital quantum fluctuation in a V-V pair.  
The molecular orbital model corresponds to the large
$h$ limit. In the opposite limit $h \rightarrow 0$, the spin and orbital
ordering depends on the relative strengths of the intraplane hoppings. In a
reasonable parameter range, we found a ground state with RS spin and
an orbital ordering corresponding to the pattern reported by
Paolasini et al\cite{paolasini}. 
However, since in this state the V ions of a vertical pair are in $(e_g e_g)$ 
and $(e_g a_{1g})$ configurations, the electronic densities are very different.
This should lead to different local distortions of the O octahedra, which 
is inconsistent with the monoclinic structure reported by Dernier and
Marezio\cite{dernier}, where all V are equivalent\cite{com3}. So we do not
think that this kind of orbital ordering is realized in V$_2$O$_3$. 

\begin{figure}[hp]
\centerline{\psfig{figure=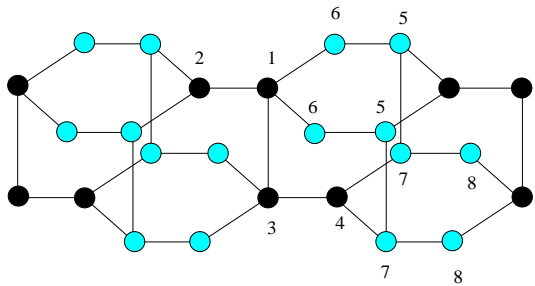,width=7.0cm,angle=0}}
\vspace{0.5cm}
\caption{The structure of the low temperature 
monoclinic phase of V$_2$O$_3$. The
gray and filled circles correspond to spin-up and spin-down orientations
of the local magnetic moments on V ions.}
\label{fig3}
\end{figure}

To summarize, we have proposed a model for the AF insulating 
phase of V$_2$O$_3$.
This model seems to be the only way to combine basic facts about the electronic
structure (S=1, orbital degeneracy, strong coupling along vertical pairs) into 
a coherent picture that agrees with all experiments. Further 
investigation of the
resulting two-fold degenerate, S=2 model for the vertical pair is in progress.

We acknowledge useful discussions with  W. Bao, B. Normand, P. M. Platzmann,
G. Sawatzky, L. H. Tjeng and C. Vettier, 
and the hospitality of the Center for Theoretical Studies of ETH
Z\"urich. This work was in part supported by the DOE grant DE/FG03-98ER45687,
Russian Foundation for Basis Reasearch grants RFFI-98-02-17275 and
RFFI-00-15-96575.


\begin{references}
\bibitem{com2}
In deriving this formula, we have not included the pair hopping term for
simplicity.  
Since this term is active only on configurations with doubly 
occupied orbitals, 
it is usually irrelevant for the low-energy physics. 
In fact we confirmed by the exact diagonalization of two V cluster 
that the energy difference coming from the pair hopping is 
less than 1\% for low-lying energy levels.
\bibitem{com3} This is
not the  case in Castellani et al's picture because, although the orbital
ordering  corresponds to the same pattern, the electronic densities are the
same at all vanadium sites.
\end{references}
\end{document}